\documentclass[%
 reprint,
superscriptaddress,
 amsmath,amssymb,
 aps,
pra,
floatfix,
]{revtex4-2}

\usepackage{graphicx}
\usepackage{physics}
\usepackage{dcolumn}
\usepackage{bm}
\usepackage[colorlinks=true, allcolors=blue]{hyperref}
\usepackage{soul}
\usepackage{outlines}
\usepackage{xcolor}
\usepackage[T1]{fontenc}
\usepackage[utf8]{inputenc}

\newcommand\revision[1]{#1}

\begin{document}

\preprint{APS/123-QED}

\title{Short two-qubit pulse sequences for exchange-only spin qubits in 2D layouts}
% \title{Exchange-only spin qubits in 2D layouts}

\author{Jason D. Chadwick}
\email{jchadwick@uchicago.edu}
\affiliation{Technology Research, Intel Corporation, Hillsboro, OR 97124}
\affiliation{Department of Computer Science, University of Chicago, Chicago, IL 60637}
\author{Gian Giacomo Guerreschi}
\author{Florian Luthi}
\author{Mateusz T. M{\k{a}}dzik}
\author{Fahd A. Mohiyaddin}
\author{Prithviraj Prabhu}
\author{Albert T. Schmitz}
\author{Andrew Litteken}
\author{Shavindra Premaratne}
\author{Nathaniel C. Bishop}
\email{nathaniel.bishop@intel.com}
\author{Anne Y. Matsuura}
\author{James S. Clarke}
\affiliation{Technology Research, Intel Corporation, Hillsboro, OR 97124}

\date{\today}

\begin{abstract}
Exchange-only (EO) spin qubits in quantum dots offer an expansive design landscape for architecting scalable device layouts. The study of two-EO-qubit operations, which involve six electrons in six quantum dots, has so far been limited to a small number of the possible configurations, and previous works lack analyses of design considerations and implications for quantum error correction. Using a simple and fast optimization method, we generate complete pulse sequences for CX, CZ, iSWAP, leakage-controlled CX, and leakage-controlled CZ two-qubit gates on 450 unique planar six-dot topologies and analyze differences in sequence length (up to 43\% reduction) across topology classes. In addition, we show that relaxing constraints on post-operation spin locations can yield further reductions in sequence length; conversely, constraining these locations in a particular way generates a CXSWAP operation with minimal additional cost over a standard CX. We integrate this pulse library into the Intel quantum stack and experimentally verify pulse sequences on a Tunnel Falls chip for different operations in a linear-connectivity device to confirm that they work as expected. Finally, we explore architectural implications of these results for quantum error correction. Our work guides hardware and software design choices for future implementations of scalable quantum dot architectures.

\end{abstract}

\maketitle

\section{Introduction}
Fault-tolerant quantum computation is expected to require a large-scale quantum computer consisting of up to tens of millions of physical qubits \cite{litinski_game_2019, gidney_how_2021, chamberland_universal_2022, kim_faulttolerant_2022, beverland_assessing_2022, leblond_realistic_2024}. Spin qubits in semiconductor quantum dots have emerged as a promising candidate for a scalable quantum architecture due to the small characteristic size of individual qubits \cite{burkard_semiconductor_2023}, long coherence times, nanosecond operation timescales, operating temperature above one kelvin \cite{vandersypen_interfacing_2017, petit_universal_2020, yang_operation_2020, camenzind_hole_2022, huang_highfidelity_2024}, and compatibility with existing fabrication techniques in the well-established semiconductor industry \cite{loss_quantum_1998, maurand_cmos_2016, veldhorst_silicon_2017, zwerver_qubits_2022, burkard_semiconductor_2023, neyens_probing_2024, george_12spinqubit_2024}. The fidelity of quantum operations on spin qubit platforms has progressed rapidly \cite{veldhorst_addressable_2014, yoneda_quantumdot_2018, west_gatebased_2019, urdampilleta_gatebased_2019, zheng_rapid_2019, noiri_fast_2022, xue_quantum_2022, mills_twoqubit_2022,  blumoff_fast_2022, madzik_precision_2022, steinacker_300_2024, huang_highfidelity_2024, tanttu_assessment_2024}, and device sizes of up to 12 have recently been demonstrated in Si/SiGe dots \cite{philips_universal_2022, weinstein_universal_2023, george_12spinqubit_2024}.

Exchange-only (EO) spin qubits, in particular, present a compelling platform for quantum computation, as they can be controlled solely through the exchange interaction between electron spins, eliminating the need for precise local magnetic field or microwave control and consequently reducing hardware complexity \cite{divincenzo_universal_2000, bacon_universal_2000}. Each EO qubit occupies three adjacent quantum dots, and one- and two-qubit operations are performed via a series of exchange interactions between pairs of dots. Recent work has experimentally demonstrated high-fidelity one- and two-EO-qubit operations \cite{weinstein_universal_2023, sun_fullpermutation_2024}.

The emergence of two-dimensional quantum dot arrays \revision{\mbox{\cite{borsoi_shared_2024, raach_fabrication_2024, ha_twodimensional_2025, john_twodimensional_2025}}} underscores the need for thoughtful layout choices in large arrays. Key considerations include integration of internal readout components \cite{oakes_fast_2023}, minimizing interconnects and control lines \cite{veldhorst_silicon_2017, li_crossbar_2018, franke_rents_2019, boter_sparse_2019, xue_cmosbased_2021}, and enabling high-speed, high-fidelity qubit operations.

In this work, we focus on the latter, investigating how the connectivity of the relevant dots within a larger array influences the performance of qubit operations. The architecture of quantum dot arrays for EO qubits presents a rich design space with numerous possibilities for arranging and connecting EO qubits. However, exploration of two-EO-qubit operations has been limited to a small subset of potential configurations \cite{shi_fast_2012, setiawan_robust_2014}. In this work, we investigate the interplay between architecture layout and two-qubit gate quality by optimizing pulse sequences directly and using the results to inform the design of layouts that admit better error correction.
 
We introduce an efficient method for mapping fixed all-to-all pulse sequences to arbitrary restricted dot connectivities, enabling the exploration of a vast array of potential device layouts. This enables us to generate pulse sequences for several two-qubit gates across 450 unique planar six-dot topologies. We find significant differences in pulse sequence length across different dot connectivities, showing that EO qubit connectivity should be considered when designing large dot arrays. We also investigate the impact of relaxing constraints on post-operation spin locations, yielding free additional reductions in sequence length with the help of the compiler. A specific choice of post-operation spin locations leads to a rapid CXSWAP operation that is almost the same cost as a standard CX. We explore the architectural implications of our findings for quantum error correction by comparing surface code performance on different dot layouts.

\revision{This optimization method allows for novel pulse sequence designs to be efficiently adapted to any qubit connectivity, enabling connectivity-agnostic pulse sequence design.} In addition, our research aims to guide both hardware and software design choices for future implementations of scalable quantum dot architectures. By providing a thorough analysis of the effects of EO qubit connectivity on two-qubit pulse sequences, we contribute to the ongoing effort to realize practical, scalable, and fault-tolerant quantum computers based on semiconductor quantum dots.

\section{Background}

\subsection{Exchange-only spin qubits}

Exchange-only spin qubits are a promising platform to scalably encode qubits in silicon quantum dots. Each EO qubit is encoded in a decoherence-free subspace (DFS) of a three-electron-spin system, distributed over three quantum dots \cite{divincenzo_universal_2000, bacon_universal_2000, fong_universal_2011}. The $\ket 0$ and $\ket 1$ states of the EO qubit correspond to the singlet and triplet states of the first two spins in the DFS, while the third spin is known as the \emph{gauge}. We refer to Ref. \cite{fong_universal_2011} for a more comprehensive explanation of the structure of the DFS.

Single-EO-qubit operations are performed via exchange interactions between component spins. Each pulse is a partial swap operation $U_{\text{SWAP}}(j,k,\theta) = \cos(\theta/2)\text{I}_{jk} + i\sin(\theta/2)\text{SWAP}_{jk}$ between two spins with some angle $\theta$ \cite{weinstein_universal_2023}. Note that an exchange pulse with $\theta=\pi$ corresponds to a complete swap (up to global phase) of the two spin states. In hardware, the exchange interaction is performed by lowering a potential barrier between two spins in adjacent quantum dots. Either the barrier voltage or the duration of the operation can be tuned to generate exchange interactions of different angles. The allowed exchange interactions between spins are thus defined by the presence of barriers between adjacent dots. In this work, we assume that the barrier voltage is the tunable knob, so that each exchange pulse is of the same duration regardless of the rotation angle.

In the encoded EO qubit picture, an exchange interaction $U_{12}(\theta)$ between spins 1 and 2 generates a Pauli Z rotation $Z(\theta)$, and interactions $U_{23}(\theta)$ and $U_{13}(\theta)$ generate rotations about two other axes in the X-Z plane of the Bloch sphere \cite{andrews_quantifying_2019}. Any two of these three exchange axes yields universal single-qubit control. A two-EO-qubit operation consists of a sequence of pairwise exchange pulses in a six-spin (two-qubit) system \cite{kempe_theory_2001, fong_universal_2011}; several pulse schedules have previously been found that implement CX and CZ operations on a small number of dot connectivities \cite{fong_universal_2011, shi_fast_2012, setiawan_robust_2014}. However, because the set of possible exchange interactions is restricted by the set of inter-dot connections (barriers) present in the hardware, not every pulse sequence can be applied as-is to a given configuration of EO qubits on a device. Existing pulse sequences can be applied to any connected configuration by inserting spin swaps (exchange interactions with angle $\pi$) on demand to bring spins near each other; however, this method can lead to extremely long pulse sequences (see Section \ref{sec:methods/generating}), which is undesirable as exchange errors accumulate linearly with pulse count \cite{andrews_quantifying_2019} and magnetic noise error scales quadratically with sequence duration \cite{weinstein_universal_2023}. In this work, we focus on generating pulse sequences with minimal pulse count with the aim of reducing both sources of error.

\begin{figure}
    \centering
    \includegraphics[width=0.9\linewidth]{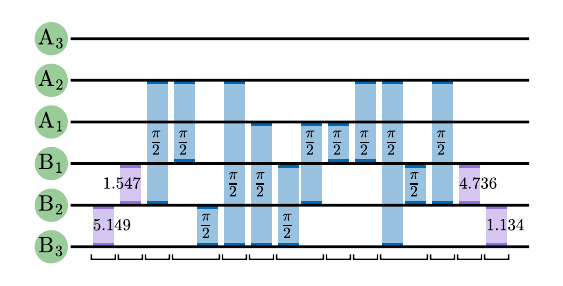}
    \caption{A 16-pulse sequence implementing a CX gate on two exchange-only qubits encoded in six spins, obtained by adding local corrections to the 12-pulse sequence from \cite{setiawan_robust_2014}. Each dot is labeled according to its qubit (A or B) and spin index, with gauge spins labeled A3 and B3. Each pulse is a partial swap operation between two spins \cite{weinstein_universal_2023}, which can be implemented with the exchange interaction \cite{divincenzo_universal_2000}. Brackets underneath show groups of sequential pulses which act on disjoint dots, and thus can in principle be applied in parallel. A different choice of local corrections can instead yield CZ; see Appendix \ref{app:all-to-all}.}
    \label{fig:a2a-cx}
\end{figure}

\subsection{Previous work on two-qubit gate sequences for EO qubits}

Ref. \cite{divincenzo_universal_2000} provided the first exchange-only two-qubit pulse sequence construction, consisting of 19 exchange pulses in 13 layers that generates a unitary locally equivalent to CX on a linear dot array (six dots connected in a line). Using a genetic evolution algorithm, Ref. \cite{fong_universal_2011} discovered a new pulse sequence (now known as the FW sequence) for a linear dot array that is valid regardless of the state of the gauge qubit. This sequence implements CX exactly and uses 22 pulses in 13 layers. Ref. \cite{shi_fast_2012} used a similar method to find CX pulse sequences for two nonlinear connectivities. Ref. \cite{setiawan_robust_2014} used a more efficient constrained search to generate \emph{locally equivalent} (correct up to single-qubit gates) pulse sequences for all-to-all connectivity and several constrained dot connectivities, recovering the FW-CX for the linear connectivity case. Figure \ref{fig:a2a-cx} shows how the 12-pulse locally equivalent CX from \cite{setiawan_robust_2014} can be extended to exactly implement CX on an all-to-all dot connectivity. Ref. \cite{zeuch_simple_2016} formalized the FW construction and Ref. \cite{zeuch_efficient_2020} generalized the construction to give sequences for CPHASE gates of arbitrary angle.

More recently, Ref. \cite{weinstein_universal_2023} demonstrated the first experimental two-EO-qubit gates, performing FW-CX and SWAP operations on six dots in a linear array with fidelities 96.3\% and 99.3\% respectively. This work also introduced the leakage-controlled CX and CZ (LCCX and LCCZ), which prevent the spread of leakage from qubit to qubit at the cost of a 65-70\% increase in pulse count (and comparable increase in interleaved randomized benchmarking error) relative to the FW-CX and CZ.

In this work, we describe and evaluate an alternative method to obtain pulse sequences for arbitrary dot connectivities, and we present pulse sequences on a much larger set of unique connectivities that are relevant to consider when designing upcoming two-dimensional dot arrays.

\subsection{Quantum error correction on spin qubits}

It is generally assumed that physical quantum hardware will never reach the algorithmically relevant error rates of $10^{-6}$ to $10^{-12}$ needed for useful applications \cite{gidney_how_2021, beverland_assessing_2022} due to unavoidable noise. To bridge the gap, quantum error correction (QEC) will be used to build a logical abstraction layer upon which high-fidelity programs can be implemented. With sufficiently-low physical error rates, QEC can exponentially suppress errors by operating with \emph{logical qubits} composed of many physical qubits that undergo repeated error detection and correction cycles. \cite{knill_resilient_1998, kitaev_faulttolerant_2003, aharonov_faulttolerant_2008}.

The surface code \cite{kitaev_quantum_1997, dennis_topological_2002} is a leading approach to quantum error correction and is attractive due to its high physical error threshold, planar connectivity requirements, and relative ease of decoding, as well as added benefits of several specialized flavors that have recently been developed to adapt to specific noise channels, basis gates, or connectivities \cite{bonillaataides_xzzx_2021, mcewen_relaxing_2023}. The fundamental action in quantum error correction is \emph{syndrome extraction}, which primarily consists of layers of parallel two-qubit gates followed by measurement of specific \emph{ancilla} qubits. Logical operations in the surface code consist of different patterns of syndrome extraction on different sets of qubits. It is important to note that, at the physical hardware level, QEC (and, more specifically, repeated syndrome extraction) is the \emph{only} algorithm we expect to run directly on the physical hardware itself. The exponential suppression of errors means that every gain (or loss) in performance at the physical level is exponentially amplified at the logical level, so it is critical to design the physical architecture to provide the best performance when running quantum error correction.

Several works have proposed novel encodings of QEC on spin qubit arrays \cite{cai_looped_2023, strikis_quantum_2023, siegel_early_2024, hetenyi_tailoring_2024}, but none have yet considered exchange-only qubits. Here, we explore simple mappings of the standard surface code and the 3-CX surface code \cite{mcewen_relaxing_2023} to several abstract hardware layouts, focusing on how differences in pulse schedules translate to differences in QEC resource estimates. A more thorough study of QEC tradeoffs on exchange-only qubits is left as future work.

\section{Methodology}

\subsection{Enumerating dot connectivities}\label{sec:methods/enumerating}

We consider a broad selection of configurations of EO qubits, which we call \emph{dot topologies}. A dot topology consists of six dots (holding two EO qubits) and a set of allowed exchange interactions, which we represent as edges connecting pairs of dots. In other words, a dot topology is a unique six-node undirected graph. We consider a total of 450 unique topologies in this work (all 6-node connected subgraphs of the square lattice, as well as several selected subgraphs of the triangular lattice), so we condense them into \emph{permutation equivalence classes} for easier visualization. An equivalence class consists of all topologies that are equivalent up to intraqubit spin permutations (changing positions of spins within each EO qubit, but maintaining the same overall qubit-level footprints). Each class thus consists of up to $3!^2 = 36$ topologies. Figure \ref{fig:greedy-lengths}a shows all permutation equivalence classes that we consider in this work.

\subsection{Generating pulse sequences}\label{sec:methods/generating}

Many methods could be used to generate pulse sequences for arbitrary connectivities. Refs. \cite{fong_universal_2011} and \cite{shi_fast_2012} used a genetic algorithm to find the FW sequence for a CX on a linear connectivity. In principle, this approach could be applied to find pulse sequences for any connectivity of interest. However, the approach is expensive (Ref. \cite{fong_universal_2011} reports a runtime of several weeks), and there is no guarantee that each optimization will converge to a desirable-length solution, which is a concern when considering a large number of different hardware connectivities. Ref. \cite{setiawan_robust_2014} uses a constrained exhaustive search method instead, and successfully finds locally-equivalent pulse sequences for several different dot connectivities. Howewer, any exhaustive method will scale poorly for operations that require longer sequence lengths, such as the leakage-controlled operations from \cite{weinstein_universal_2023}; additionally, the constraints placed on the optimization (time-symmetric sequences with pulse angles only allowed to be multiples of $\pi/2$) restrict the possibilities and require that the optimization can only be done up to local corrections, not allowing the pre- and post-sequence single-qubit corrections to be incorporated into the optimization.

Instead, inspired by the qubit routing problem in quantum compiler research \cite{li_tackling_2019, nannicini_optimal_2021}, we begin with a pulse sequence designed for all-to-all dot connectivity and aim to find an efficient set of spin-swaps (exchange pulses of angle $\pi$) to map this sequence to any restricted connectivity of interest. For example, consider a linear connectivity with spins assigned $A_3-A_2-A_1-B_1-B_2-B_3$. Suppose we wish to apply the pulse pair $(A_2, B_2)$, but these two spins are not directly connected. We can insert the spin swaps $(A_2, A_1)$ and $(A_1, B_1)$ to bring the spin originally in $A_2$ to the location of $B_1$, where it can now interact with $B_2$. 

Swap-adding is motivated by the observation that every known exchange-only CX sequence fundamentally aligns with the theory of the FW sequence \cite{zeuch_simple_2016}, which involves three ``quasi-Fredkin'' \cite{weinstein_universal_2023} operations on sets of the component spins. We therefore attempt to efficiently map these fundamental operations onto each connectivity rather than solving the problem anew each time.

\begin{figure}
    \centering
    \includegraphics[width=\linewidth]{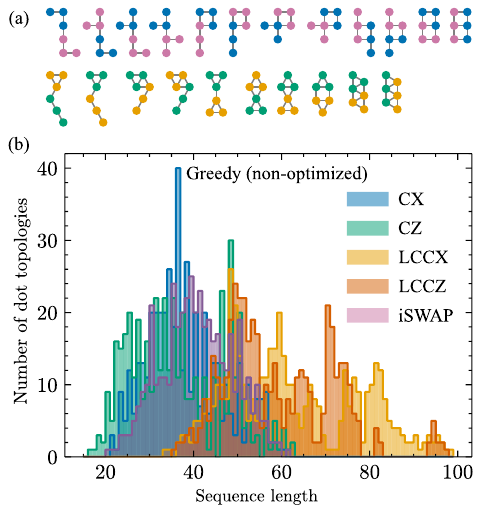}
    \caption{(a) All permutation classes of dot topologies considered in this work. Dots of the same color indicate that they hold spins of the same EO qubit. Top row contains all topologies that can arise in a 2D rectangular grid of dots. Bottom row consists of additional topologies that require triangular EO qubits. (b) Starting with an all-to-all pulse sequence and greedily inserting spin swaps to adapt it to a restricted dot topologies yields poor results. Each colored histogram shows the distribution of resulting sequence lengths across all 450 dot topologies for a specific two-qubit operation. See Appendix \ref{app:all-to-all} for initial all-to-all pulse sequences.}
    \label{fig:greedy-lengths}
\end{figure}

There are several practical advantages to the swap-adding approach. First, compared to the constrained search method used by Ref. \cite{setiawan_robust_2014}, we no longer need to work under a symmetry constraint. Second, our swap-adding optimization runs very quickly; on average, it takes seconds or a few minutes to optimize a pulse sequence for a single topology, and parallelization across different topologies and different gate operations is trivial. This improvement in runtime compared to earlier works makes it far easier to iterate with new optimization goals and could even enable more fine-grained calibration such as generating unique hardware-aware pulse sequences for different quantum dots on a device. Finally, it is easy to find a valid solution for any connectivity, so we will never be left without any solution; for example, the simplest swap-adding algorithm is to step through the all-to-all sequence and greedily add spin swaps to enable each new pulse. However, the greedy swap-adding approach mentioned previously is highly suboptimal, as we show in Figure \ref{fig:greedy-lengths}b; this approach yields sequence lengths of up to 60 for a CX, depending on the topology, while we expect sequence lengths to be 28 at worst (existing linear connectivity pulse sequence). This motivates a more intelligent approach.

\begin{figure*}
    \centering
    \includegraphics[width=\linewidth]{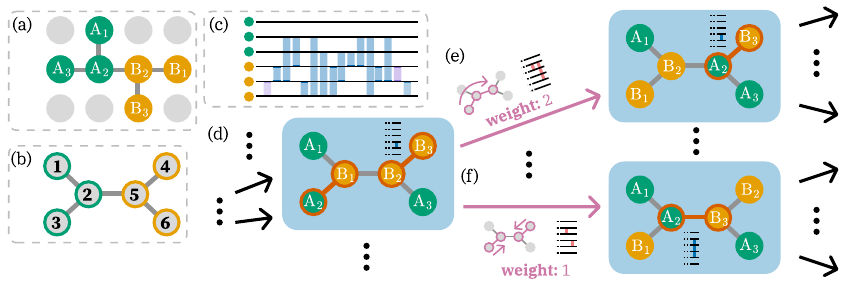}
    \caption{The optimization graph used to find a pulse sequence for a given dot configuration, described in Section \ref{sec:methods/sequences/graph}. (a) Configuration of the two EO qubits of interest within a larger 2D grid of dots, with labeled spins comprising qubits A and B. Allowed exchange interactions are indicated by lines between dots. (b) Connectivity of the selected (numbered) dots, encoding the allowed exchange interactions. (c) Ideal \emph{all-to-all} CX pulse sequence (assuming any exchange interaction is allowed). Not all of these edges are present in the connectivity of interest, so spin swaps must be interleaved to shuffle the spins between dots and bring the desired spins into contact. (d) A node of the optimization graph. Each node corresponds to the application of one of the reference pulses from the all-to-all sequence. At this layer, the pair $(B_2, B_3)$ is applied, so $B_2$ and $B_3$ must be in neighboring dots. Note the highlighted dot pairs $(A_2, B_1)$ and $(B_2, B_3)$ in red, corresponding to the most-recently-applied pulses. (e) An edge between nodes represents the insertion of spin-swap operations (exchange interactions). In this case, we perform spin swaps to move spin $A_2$ from dot 3 to dot 5, where it can interact with $B_3$ as desired. The weight of this edge is the \emph{total number of new pulses} that will be added to the sequence: the first spin swap on dots (2,3) can be absorbed the previous pulse $(A_2, B_1)$; however, the second spin swap and the $(A_2, B_3)$ reference pulse each add one to the sequence length, so the overall edge weight is 2. (f) An edge to a different node in the next layer. In this case, we apply two spin swaps which can both be absorbed into previous pulses. The weight is 1 due to the $(A_2, B_3)$ pulse that is applied in the next node.}
    \label{fig:optimization}
\end{figure*}

\subsubsection{Formulating swap routing as a shortest path problem}\label{sec:methods/sequences/graph}

There are two important considerations when determining which spin swaps to insert to map a sequence to a dot topology. First, two exchange pulses that act in succession on the same dot pair can be \emph{merged} into one pulse with rotation angle equal to the sum of the original two; in effect, on a hardware where exchange pulses are fixed duration, the second pulse has become ``free''. When incrementally constructing a sequence, we say that a previous exchange pulse is \emph{unblocked} if neither dot has participated in a different exchange pulse since then (meaning that if we were to apply a new pulse on the same pair, we could absorb it into this previous pulse). Second, when two spins need to interact but are not currently in adjacent dots, there are often several choices of swap sequences that will bring them into contact. The choice of swap sequence can either help or hinder later operations in the pulse sequence, depending on where all six spins end after the application of the swaps. A no-lookahead greedy approach will rarely choose the best intermediate swap sequence at every step.

We encode the swap-adding optimization problem as a shortest-path problem on a layered directed acyclic graph, where each path from the source node to the destination node encodes a valid sequence of exchange pulses that realizes the desired gate. We begin by specifying the target dot connectivity (Figure \ref{fig:optimization}a-b) and the \emph{reference sequence}, the all-to-all pulse sequence that implements the desired operation (Figure \ref{fig:optimization}c). Each layer of the optimization graph corresponds to one \emph{reference pulse} from this sequence. Within each layer are many nodes, each representing the application of the reference pulse for a different configuration of spins within the dots (Figure \ref{fig:optimization}d-f). A configuration is only valid for a given layer if the two spins involved in the corresponding reference pulse are in adjacent dots, guaranteeing that the reference pulse can be applied. Each node has outgoing directed edges that lead to some nodes in the next layer, with edge weights related to the number of spin swaps (exchange pulses of angle $\pi$) needed to transition from the first node to the second by permuting the spins to the new configuration. Figure \ref{fig:optimization} shows several nodes of an example optimization graph.

Each node also has information on the current ``unblocked exchange pulses'' that are available to absorb a potential future pulse. The weight of an edge is the total number of additional pulses required to perform the specified spin swaps \emph{and} the destination reference pulse, after absorbing any possible pulses into previous unblocked pulses. 

Finally, there is a single source node that specifies the initial locations of all six spins in the dot topology, and there is a single destination node which does not correspond to a reference pulse but instead enforces that the final locations of all six spins are identical to the initial locations. In this optimization graph, any path from source to destination must contain one and only one node from each layer, and so the sum of edge weights along any path from source to destination is equal to the length of the pulse sequence formed by the pulses on the nodes and edges of the path. We can therefore use a simple shortest-path algorithm to find the shortest sequence. The exact pulses of the sequence can then be constructed by traversing this path and inserting the specified swaps and reference pulses from each edge and node along the way.

Our code is written in Python using the \texttt{networkx} package. In the problem instances that we consider, we find that we can exactly solve any single swap-adding optimization in only a few minutes. However, for larger problems, such as a single-step weight-four parity check \cite{reagor_hardware_2022}, the search graph could be constructed on demand instead of ahead of time, and approximate or heuristic methods could be used for pathfinding.

\subsubsection{Correctness and optimality}
\revision{The goal of the proposed optimization method is to schedule the pulses in the same order as they appear in the all-to-all sequence, but satisfying the connectivity constraints between dots by inserting spin swaps as needed. We now explain why our method finds the \emph{optimal} (shortest sequence length) such schedule.}

\revision{Each node layer of the graph contains every possible spin configuration in which the two spins that need to interact are in adjacent dots. Between two adjacent layers, we include a forward edge between every pair of nodes corresponding to the shortest sequence of spin swaps that properly permutes the spin configuration. Because each node represents the application of a reference pulse in a specific spin configuration, and every edge is a valid transition between nodes, any path through the search graph represents a sequence composed of spin swaps and reference pulses. Therefore, because every reference pulse is applied to the correct pair of spins in the correct order and the initial and final configurations are guaranteed by the definition of the source and destination node, any path from the source node to the destination node \emph{correctly} implements the desired operation.}

\revision{Because we include every possible valid node and edge at each layer, we ensure that every unique pulse sequence that can be obtained by inserting spin-swaps into the reference pulse is represented by some path through this graph (except for sequences for which an equivalent shorter sequence exists by replacing a specific swap-edge with a shorter equivalent swap-edge). Therefore, the shortest solution sequence corresponds to one of these paths. Consequently, because the sum of edge weights along a path is exactly equal to the length of the corresponding pulse sequence by construction, the shortest solution sequence will be represented by the shortest path through the search graph. We can therefore claim that this graph-based search will find the \emph{optimal} solution to map a fixed all-to-all sequence to a specified dot topology by finding the shortest path.}

However, we note that this optimization method does not currently account for the flexibility of changing the order of certain pulses in the all-to-all sequence. For example, the fourth and fifth pulses in the all-to-all CX shown in Figure \ref{fig:a2a-cx} can be applied in either order, but our current optimizer does not directly account for this. We experimented with changing the ordering of pulses in the reference sequence and observed slight variations (1-2 pulses) in sequence lengths for some specific connectivities, but did not observe any systematic shifts in aggregated sequence length results. We intend to extend the optimizer to account for this flexibility in the future.

\begin{figure*}
    \centering
    \includegraphics[width=\linewidth]{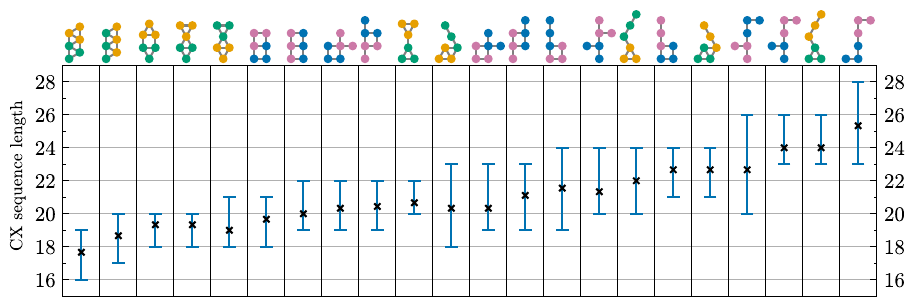}
    \caption{Optimized CX sequence lengths, grouped by permutation equivalence class. Within each class, blue line spreads from minimum to maximum sequence length and black x denotes mean. For rectangular-grid topologies, dark blue represents the ``control'' qubit of the CX and magenta the ``target''. For non-rectangular-grid topologies, green represents the ``control'' and gold the ``target''.}
    \label{fig:all-lengths-cx}
\end{figure*}

\subsection{Modifying final spin locations}\label{sec:methods/permutations}

We now discuss two examples of modifications to the search graph that can yield interesting alternative operations. In the implementation described previously, there is a single destination node in the search graph that corresponds to the state where all six spins are located in the dots where they began. This ensures that any valid pulse sequence must return the spins to their initial locations. However, we can change this final destination node or add other possible destination nodes to specify other optimization targets.

\subsubsection{CXSWAP}\label{sec:cxswap}

Alternatively, we can set the mapping so that spins $A_1$ and $B_1$ are exchanged, $A_2$ and $B_2$ are exchanged, and $A_3$ and $B_3$ are exchanged. This corresponds to a qubit-level SWAP operation in addition to the original reference operation. Using this optimization graph, the resulting pulse sequences will yield CXSWAP operations instead of CX, potentially with shorter pulse sequences than those required to do the distinct operations sequentially, by permuting the qubits \emph{during} the application of the operation. This yields the following two-qubit gate matrix:
\newenvironment{spmatrix}[1]
 {\def\mysubscript{#1}\mathop\bgroup\begin{pmatrix}}
 {\end{pmatrix}\egroup_{\textstyle\mathstrut\mysubscript}}
\begin{equation}\label{eq:cxswap}
    \begin{spmatrix}{\text{SWAP}}
        1 & 0 & 0 & 0 \\
        0 & 0 & 1 & 0 \\
        0 & 1 & 0 & 0 \\
        0 & 0 & 0 & 1
    \end{spmatrix}
    \cdot
    \begin{spmatrix}{\text{CX}}
        1 & 0 & 0 & 0 \\
        0 & 1 & 0 & 0 \\
        0 & 0 & 0 & 1 \\
        0 & 0 & 1 & 0
    \end{spmatrix}
    =
    \begin{spmatrix}{\text{CXSWAP}}
        1 & 0 & 0 & 0 \\
        0 & 0 & 0 & 1 \\
        0 & 1 & 0 & 0 \\
        0 & 0 & 1 & 0
    \end{spmatrix}
\end{equation}

CXSWAP-like gates are the fourth and final class (up to single-qubit Cliffords) of Clifford two-qubit operations along with Identity-like, SWAP-like, and CX-like gates. Consequently, a fast CXSWAP operation provides interesting opportunities for QEC design, where syndrome extraction circuits are typically made up entirely of Clifford gates. Ref. \cite{mcewen_relaxing_2023} showed that the surface code can be expressed with CXSWAP (or iSWAP, which is locally equivalent) as the two-qubit basis gate rather than CX, suggesting that these gates may be useful for other QEC circuits. For example, by replacing some CXs with CXSWAPs, a 4+1 flag fault-tolerant code \cite{chao_flag_2020} could be adapted to degree-4 connectivity, or the stepping surface code \cite{mcewen_relaxing_2023} could be performed in place (maintaining its advantageous leakage reduction properties while eliminating the extra qubit overhead typically needed for this circuit). A fast CXSWAP in place of a CX may also prove useful when adapting QEC to sparsely connected qubit arrays that may be necessary in spin qubits due to the relatively large size of current state-of-the-art readout components \cite{oakes_fast_2023}.

\subsubsection{Relaxing spin location constraints}

Alternatively, instead of modifying the default spin location constraint in the final layer, we can relax the constraint by allowing \emph{several} possible spin permutations in the final layer. In the search graph, we add several additional destination nodes, each of which corresponds to a different final mapping of spins. We can then search for the shortest path from the source to \emph{any} destination node (which is not significantly harder than the original shortest path problem). We find that this relaxation can lead to sequences that are several pulses shorter.

We specifically focus on the case where we allow any \emph{intraqubit} spin permutations such that the qubit-level footprints remain the same but spins 1, 2, and 3 can be shuffled within each qubit. Each optimized pulse sequence must choose a specific spin permutation to apply. When compiling a quantum program to the device, the compiler can simply track the known repositioning of spins as these operations are successively applied, and pick the correct next pulse sequence given the prior spin locations.

\begin{figure*}
    \centering
    \includegraphics[width=\linewidth]{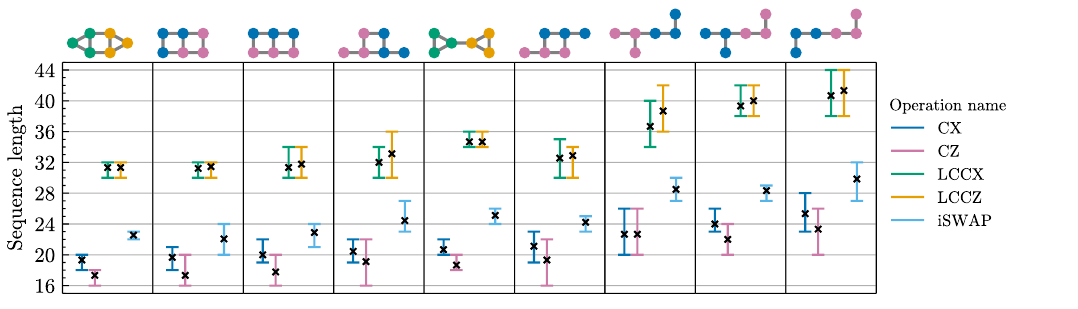}
    \caption{Optimized sequence lengths for five different two-qubit operations, for a selected set of permutation equivalence classes.}
    \label{fig:selected-lengths-by-gate}
\end{figure*}

\begin{figure*}
    \centering
    \includegraphics[width=\linewidth]{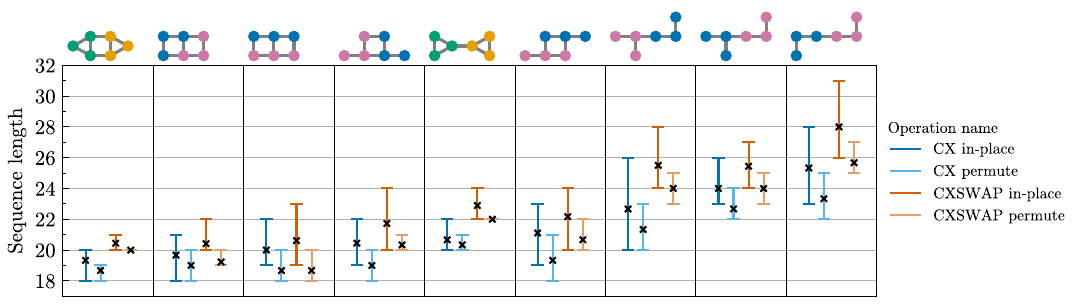}
    \caption{CX and CXSWAP sequence lengths for selected permutation equivalence classes, with and without intraqubit spin permutations allowed. Across all topologies, CXSWAP is 22.2\% shorter than applying CX and SWAP operations sequentially and is only 7.34\% longer on average than CX. For both CX and CXSWAP, allowing permutations reduces the average and maximum sequence length within each equivalence class.}
    \label{fig:selected-lengths-by-permutation}
\end{figure*}

\section{Optimized sequence results}

We run our optimizations over the full collection of dot topologies and present the results aggregated within each permutation equivalence class. \revision{We optimize for five two-qubit operations: CX, CZ, iSWAP, LCCX, and LCCZ. LCCX and LCCZ are the ``leakage-controlled'' variants of CX and CZ \mbox{\cite{weinstein_universal_2023}}, which are constructed by applying two CX-like sequences with a single-qubit operation in between, so the all-to-all reference sequences (and thus our optimized sequences) are significantly longer. The choice of all-to-all sequences used for these optimizations is described in Appendix \mbox{\ref{app:all-to-all}}.}

\subsection{Standard operation sequence lengths for different topology classes}

Figure \ref{fig:all-lengths-cx} shows the optimized sequence lengths for a CX operation on all 450 dot topologies in consideration. The results are organized by permutation equivalence class (see Section \ref{sec:methods/enumerating}). Observe that the linear dot topology class (rightmost) yields the longest pulse sequences. Generally, as the dots become more connected, shorter CX sequences become possible. For example, the ``linear-parallel'' permutation class (seventh from left) has a maximum-in-class length of 22 compared to 28 for the fully linear class, corresponding to a reduction of 21.4\%. For the furthest left topology class consisting of two densely connected triangular qubits, maximum-in-class sequence length is reduced to 19, a 32\% reduction.

Figure \ref{fig:selected-lengths-by-gate} shows optimized sequence lengths for an extended set of gates (CX, CZ, iSWAP, LCCX, and LCCZ) over a selected subset of relevant topology classes (the connectivities that we believe to be the most promising when considering integration into a larger array).  Across all five two-qubit operations, a similar trend emerges: denser dot connectivities yield shorter two-qubit operations. On average, our optimized leakage-controlled variants of CX and CZ (LCCX and LCCZ) are 60-70\% longer than the standard CX and CZ.

\begin{figure*}
    \centering
    \includegraphics[width=\linewidth]{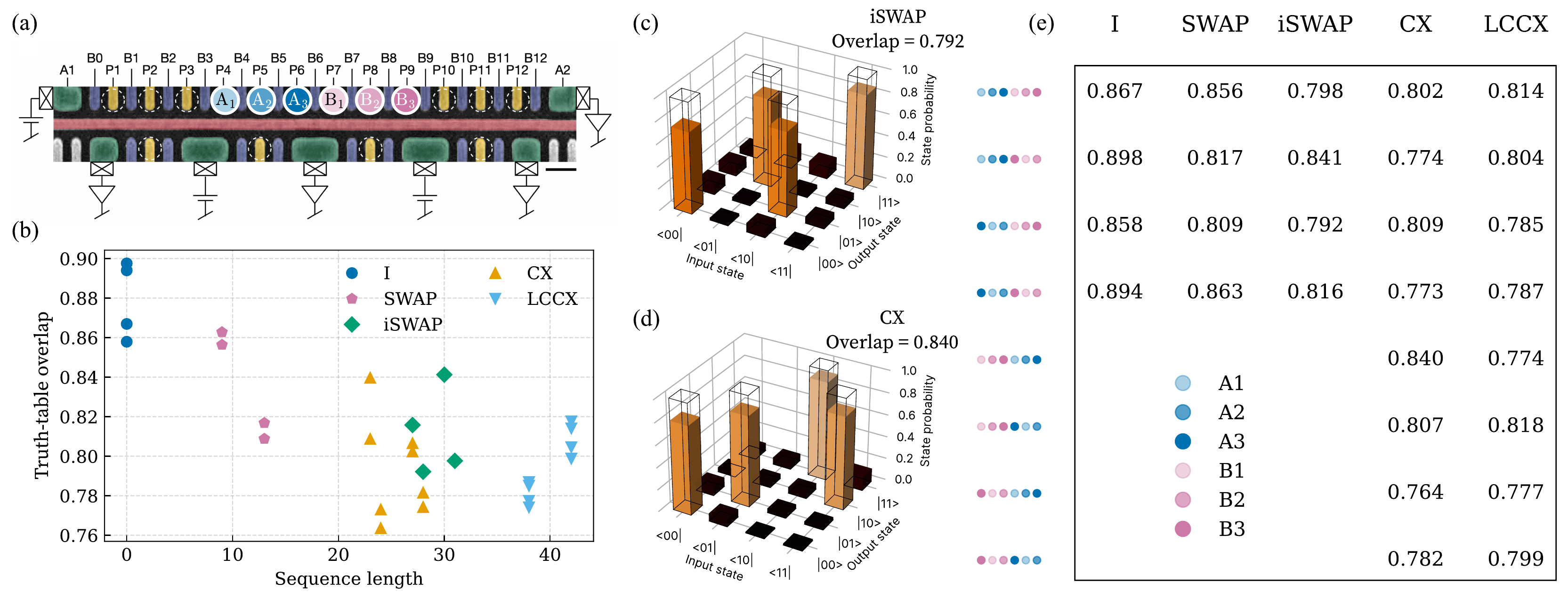}
    \caption{Experimental truth-table validation of pulse sequences. (a) 12 quantum dot Tunnel Falls device. Dots P4,P5,P6 and P7,P8,P9 are used for the two EO qubits. (b) Summarized results comparing sequence length and truth-table overlap. (c-d) Example truth tables for iSWAP and CX two-qubit gates for specific spin configurations. (e) Truth-table overlap results for the five gate types and eight spin configurations, without SPAM corrections. CX-type gates are performed with qubit A as control and qubit B as target.}
    \label{fig:experiments}
\end{figure*}

These results show the importance of considering dot connectivity when designing a large quantum dot array for exchange-only qubits. Significant improvements in two-qubit operation pulse count can be achieved simply by changing the way two EO qubits are connected. As a further example, if SWAP operations are expected to be necessary during computation, consider that a ``linear-parallel'' pair of qubits can be fully swapped with 3 exchange pulses in a single maximally-parallel step, while a linearly-connected pair of qubits will require 9 exchange pulses in 6 steps.

Finally, we note that this particular optimization did not find a 22-pulse CX for any linearly connected configuration of EO qubits, although such a sequence is known to exist \cite{fong_universal_2011} for a particular configuration. This is due to the choice of the initial all-to-all sequence (i.e. the choice of the local corrections at the start and end which are needed to generate the full CX); choosing a sequence equivalent to that from \cite{fong_universal_2011} would indeed result in our optimizer finding a 22-pulse sequence for that connectivity. For simplicity, the results we present here are all obtained from a single all-to-all CX sequence (Figure \ref{fig:a2a-cx}), but for a more heavily optimized pulse library, our optimizer could be run with many variations of initial sequences and the best result for each topology could be kept, which is still feasible due to the fast runtime of our optimizer. 

\subsection{Effects of spin permutation constraints}

Figure \ref{fig:selected-lengths-by-permutation} shows optimized sequence lengths for variants of CX with different final spin location constraints (explained in Section \ref{sec:methods/permutations}). Across all 450 dot topologies, allowing intraqubit spin permutations (``CX-permute'') yields a 0-13\% (mean 5.7\%) reduction in sequence length. Notably, the worst linear topology CX can be performed in 25 pulses instead of 28. We emphasize that there is no penalty to replacing CX with CX-permute; the compiler simply needs to track the applied permutations and account for them in future operations.

Additionally, observe that the CXSWAP operation (which is a CX sequence with final spin location constraints that enforce a qubit SWAP) is similar in length to the CX itself. Across all topologies, CXSWAP is between 8.7\% \emph{shorter} and 27\% longer (mean 7.34\% longer) compared to a standard CX. Compared to a standard CX and a standard SWAP applied sequentially, the optimized CXSWAP is 4.8-35.7\% shorter (mean 22.2\%). When intraqubit permutations are allowed, the CXSWAP-permute is an average of 7.2\% longer than the CX-permute and 23.5\% shorter than CX-permute and SWAP in series. With these optimized sequences, a SWAP operation can be applied for very low cost whenever a CX (or any other two-qubit operation) is being applied to a pair of qubits, which may be beneficial for certain QEC applications, as discussed in Section \ref{sec:cxswap}.

\section{Experimental validation}

The pulse sequences presented in the previous section have been compiled into a library available at \cite{chadwick_data_2024}. This library has been integrated into the Intel Quantum Compiler and the Intel quantum hardware stack, allowing the pulse sequences to be queried on demand based on the target dot connectivity.

This allows us to experimentally demonstrate the described pulse sequences for the CX, LCCX, iSWAP and SWAP gates. 
We use an Intel-fabricated Tunnel Falls device \cite{neyens_probing_2024, george_12spinqubit_2024} to bring a linear array of six quantum dots into the (1,3,1,1,3,1) electron regime, and tune the exchange couplings between the quantum dots so that we can encode two exchange-only qubits \cite{andrews_quantifying_2019, weinstein_universal_2023}. 
The Pauli spin blockade (PSB) technique \cite{burkard_semiconductor_2023} is used to perform the spin-to-charge conversion, allowing both qubits to be read out using a sensing quantum dot.
With PSB and post-selection, we avoid initialization into leakage states and can reliably initialize the system into its $\ket{00}$ state.

Figure \ref{fig:experiments} summarizes the experiments used to verify the gate action of the two-qubit gates of interest.
Depending on the positions of the gauge dots in the two qubits, different pulse sequences are required to perform the desired gate action. 
In the linear dot connectivity, we query pulse sequences for eight different gauge spin configurations for the CX and LCCX gates and four different configurations for the SWAP and iSWAP gates (as they act symmetrically on the two qubits).
The EO qubits are initialized in a fixed configuration; to efficiently access the different configurations, the spin states of the electrons on the respective qubits are selectively swapped by applying a series of three $\pi$-pulses after initialization.
The gate action characterization is performed using truth-table measurements in which the computational basis states are prepared, the gate under question is applied, and the resulting two-qubit state is measured in the computational basis \cite{he_twoqubit_2019, sigillito_coherent_2019, madzik_precision_2022}. We calculate the \emph{truth-table overlap} of the measured classical probabilities $P_{\text{meas}}$ with the probabilities of the ideal gate action $P_{\text{ideal}}$ over the basis states as $\text{Tr}(P_{\text{meas}} P_{\text{ideal}}^{-1})/2^{n_{\text{qubits}}}$.
State preparation and measurement (SPAM) error contributions can be characterized using the above scheme with an identity gate on both qubits, which yields an average truth-table overlap of 87.9\% for the four different configurations.
Without SPAM corrections, we extract the average truth-table overlap over the respective configurations for the SWAP, iSWAP, CX, and LCCX to be 83.6\%, 81.2\%, 79.4\%, and 79.5\% respectively. \revision{Note that these truth-table overlap values are not directly comparable to other fidelity metrics that can be measured via randomized benchmarking or process tomography, as truth-table overlap only considers the Z basis states and does not correct for SPAM. }
These results confirm that the pulse sequences work as expected; a full characterization of the gate action using quantum process tomography or similar techniques is beyond the scope of this work.

Figure \ref{fig:experiments}b shows that truth-table overlap tends to decrease as pulse sequence length increases; however, the LCCX appears to exhibit some degree of noise resilience compared to the standard CX, achieving similar error rates despite much longer sequence length. \revision{This increased error resilience of LC-type operations is consistent with what was observed in Ref. \mbox{\cite{weinstein_universal_2023}}, and corresponds to an improved insensitivity to low-frequency magnetic noise that is inherent to the underlying pulse sequence.}

\section{Case study: error correction on different hardware layouts}

\begin{figure*}
    \centering
    \includegraphics[width=\linewidth]{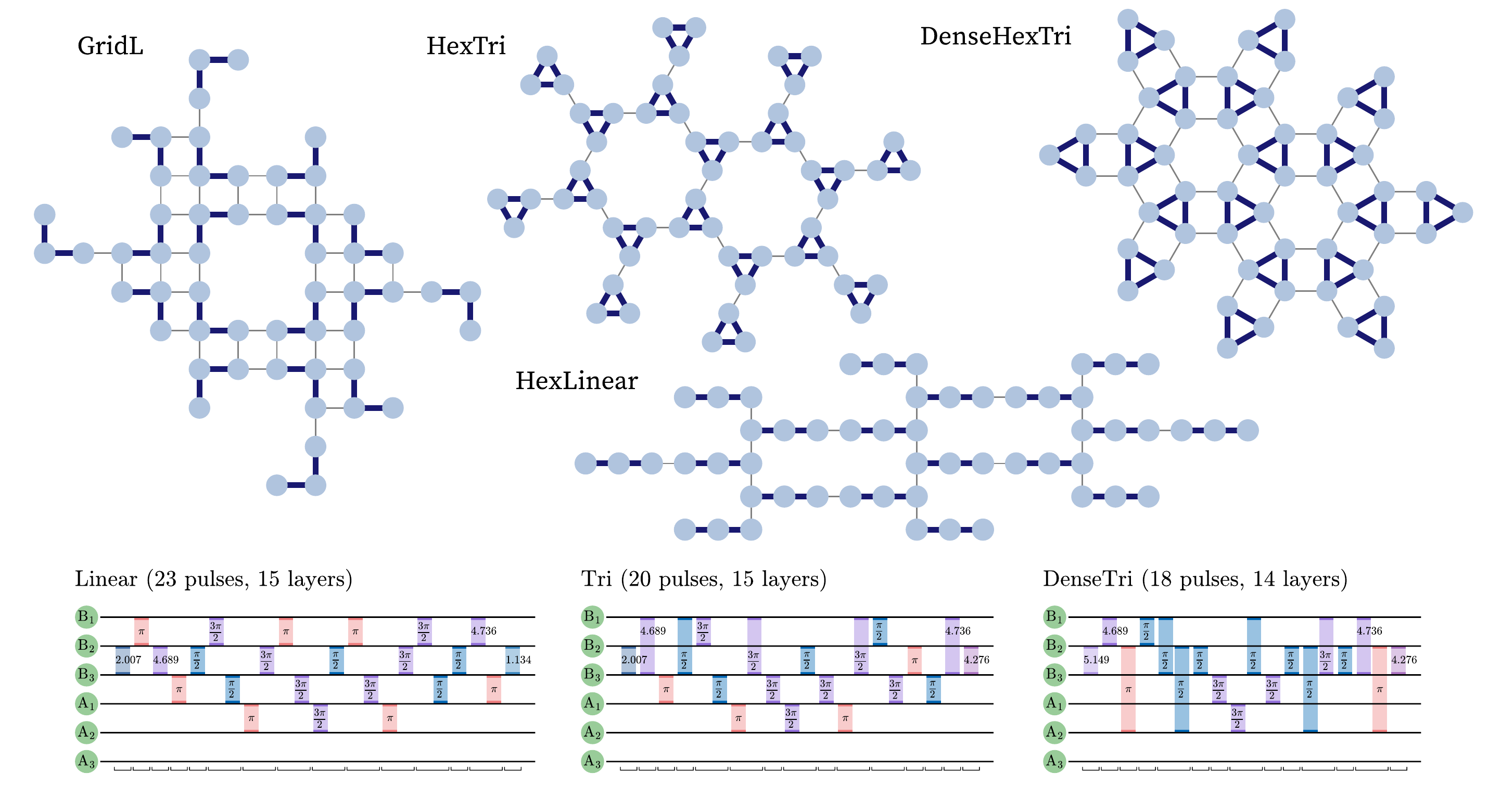}
    \caption{\emph{Top:} The four dot layouts considered in this study. Names denote the qubit-level connectivity (Grid or Hex) and the dot-level connectivity (Linear, L, Tri). Thick dark blue lines indicate intraqubit connections and thin gray lines indicate inter-qubit connections. \emph{Bottom:} Example CX pulse sequences for three different two-EO-qubit connectivities. More densely-connected qubits yield shorter pulse counts (23 for Linear vs. 18 for DenseTri); however, note that the number of fully-parallel layers (14-15) remains similar.}
    \label{fig:layouts}
\end{figure*}

While we have shown that changing EO qubit connectivity can significantly reduce pulse count for two-qubit gates, it remains unclear how this will translate to system-level, error-corrected device performance. In this section, we present several different abstract hardware layouts with different EO qubit topologies and evaluate the number of qubits needed to reach the teraquop regime ($10^{-12}$ logical error rate) in different noise regimes.

\subsection{Evaluated dot layouts}

Figure \ref{fig:layouts} shows the four layouts that we consider in this study. Layout name denotes the qubit-level connectivity (Grid or Hex) and the dot-level connectivity (L, Tri, Linear). Depending on the qubit connectivity, we map a section of the bulk of either the 4CX (grid) or 3CX (hex) surface code \cite{mcewen_relaxing_2023} to the layout. We then specify the schedule of two-qubit gates that corresponds to one round of syndrome extraction and compile the corresponding pulse-level schedule consisting of pairwise exchange pulses. Each schedule consists of four layers of CX gates. We allow post-operation intraqubit spin permutations to reduce pulse count, tracking the positions of spins during the schedule to apply the correct pulse sequence at each step. Figure \ref{fig:layout-lengths} shows the average number of pulses per CX for a QEC cycle on each layout. As expected, the linear-connectivity layout has the highest average length, and the dense triangular connectivity yields the lowest average length.

\revision{For this case study, we consider the standard CX operation because it aligns with the standard formulation of surface code circuits and allows us to more easily compare against other implementations. We do not consider the leakage-controlled variant (LCCX) because its primary benefit is in managing leakage out of the computational basis, a mechanism which does not have built-in support in the industry-standard QEC simulation tool Stim \mbox{\cite{gidney_stim_2021}}. To thoroughly investigate the effect that this has on quantum error correction is beyond the scope of this work, as it would involve significant modifications to the simulator and the decoder \mbox{\cite{googlequantumai_suppressing_2023}}. We believe an exploration of the impact and mitigation of leakage on EO qubits in the context of QEC is important future work deserving of a more in-depth study.}

\subsection{Parallelism restrictions}

Current EO experiments \cite{weinstein_universal_2023} have used fully-sequential pulse sequences, meaning that the duration of a sequence is directly determined by its pulse count. However, some degree of parallelism across a device will be necessary for large-scale quantum computation. In a two-dimensional quantum dot array, considerations such as classical control signal crosstalk \cite{weinstein_universal_2023} or gate virtualization \cite{baart_singlespin_2016, mills_shuttling_2019, volk_loading_2019, che_fast_2024} may limit the parallelism of exchange pulses. Pulse parallelism is important to reduce the gate duration $t_{\text{gate}}$; since $T_2^*$ errors scale with $(t_{\text{gate}}/T_2^*)^2$, reducing the gate duration can significantly improve performance. We consider three conditions: \emph{Full parallelism}, where each pulse in the schedule is performed as early as possible; \emph{N-restricted}, where two neighboring quantum dots cannot both perform exchange interactions with other dots at the same time; and \emph{NN-restricted}, where this restriction extends to next-nearest neighbors. We compile the pulse schedule for all CX gates in a full QEC round, scheduling each individual pulse as early as possible within the parallelism restrictions.

Figure \ref{fig:parallel-lengths} shows the total durations of the compiled QEC cycle schedules for each layout under the three parallelism conditions. Interestingly, with full parallelism, all layouts have almost identical durations, so we would expect them to all have similar levels $T_2^*$-induced errors. However, with parallelism restrictions, the HexTri layout becomes significantly shorter than the others. The DenseHexTri layout, which has the shortest pulses-per-CX count, performs relatively poorly in restricted parallelism settings, implying that denser connectivities may suffer more from neighbor-based parallelism restrictions.

From these full-QEC-cycle pulse schedules, we then recover the schedule for each individual two-qubit gate, preserving the parallelism or serialization of pulses and including any idle time induced by parallelism restrictions, to simulate individually under noise.

\begin{figure}
    \centering
    \includegraphics[width=\linewidth]{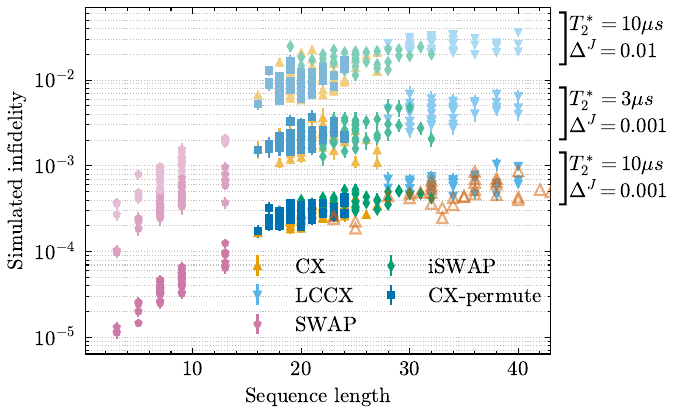}
    \caption{\revision{Under three settings of fixed noise parameters $T_2^*$ and $\Delta^J$, simulated gate infidelity scales with increasing sequence length. Unfilled orange-outlined triangles correspond to CX sequences optimized with the ``greedy'' approach (Figure \mbox{\ref{fig:greedy-lengths}}), which have a geometric mean of 1.74$\times$ increased sequence length and 1.91$\times$ increased error compared to the optimized CX sequences (blue squares).}}
    \label{fig:length_vs_infid}
\end{figure}

\begin{figure}
    \centering
    \includegraphics[width=0.7\linewidth]{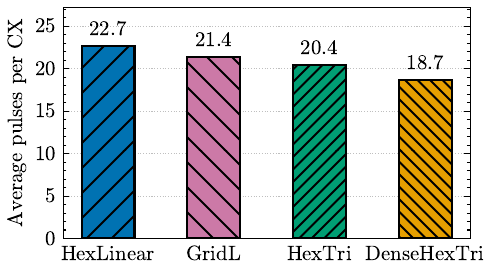}
    \caption{Average pulse count per CX for QEC round on each layout. As expected, more densely-connected layouts yield shorter pulse sequences.}
    \label{fig:layout-lengths}
\end{figure}

\begin{figure}
    \centering
    \includegraphics[width=\linewidth]{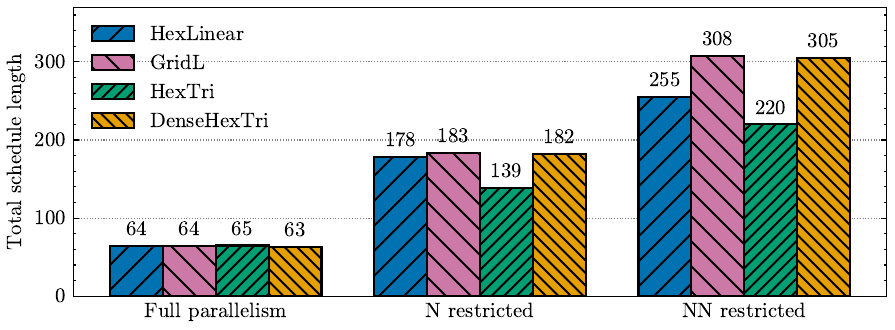}
    \caption{Schedule length for one full QEC cycle of CX gates, under various parallelism restrictions. Here, ``length'' refers to the duration of the schedule in terms of the number of steps, where many pulses can be applied in parallel during each step.}
    \label{fig:parallel-lengths}
\end{figure}

\begin{figure*}
    \centering
    \includegraphics[width=\linewidth]{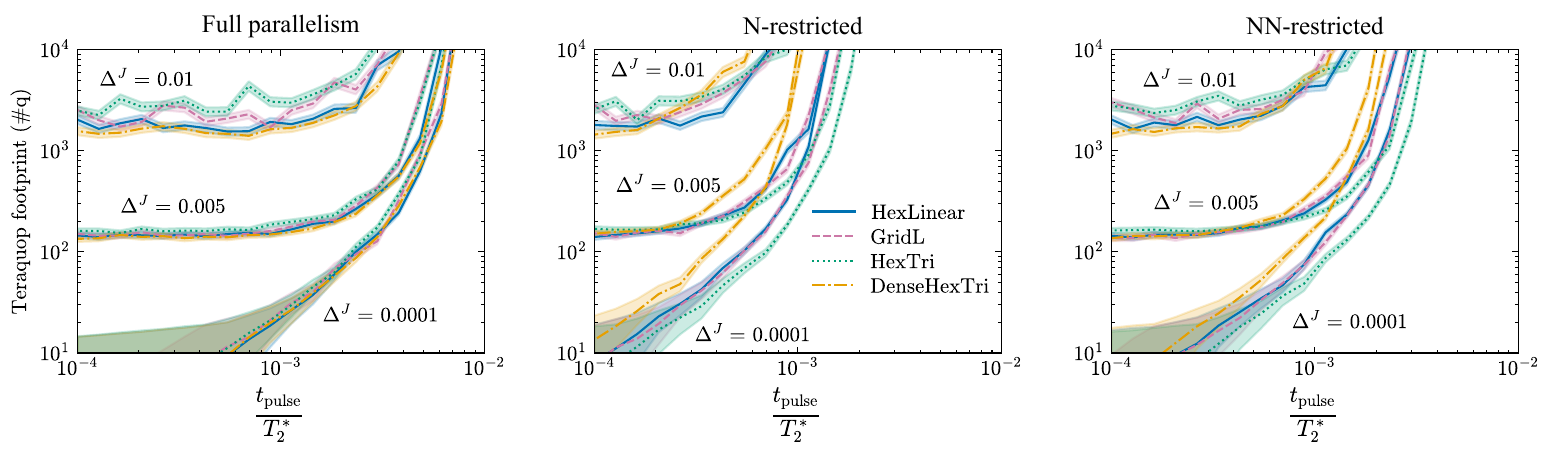}
    \caption{Teraquop footprints for all four layouts, under different parallelism conditions and in different noise regimes. The ratio $\frac{t_{\text{pulse}}}{T_2^*}$ varies along the x axis and each plot contains three groups of lines corresponding to different $\Delta^J$ strengths.}
    \label{fig:tqfps}
\end{figure*}

\subsection{Noisy simulation}

We consider noise characterized by $T_2^*$, the ergodic dephasing time of an individual spin, and by $\Delta^J$, the unitless exchange error. For fixed values of these parameters, we simulate evolution under the time-varying Hamiltonian
\begin{equation}
\begin{alignedat}{3}
    H(t) &= \epsilon^z \sum_i S_i^z + \sum_{ij} J_{ij}(t) \mathbf{S}_i \cdot \mathbf{S}_j &&\bigg\} H_{\text{ideal}}\\
    &+ \sum_i \delta_i^{\epsilon^z} S_i^z + \sum_{ij} \delta^{J}_{ij} J_{ij}(t) \mathbf{S}_i \cdot \mathbf{S}_j &&\bigg\} H_{\text{noise}},
\end{alignedat}
\end{equation}
where $\epsilon^z$ is the constant Zeeman term (assuming no magnetic field gradients or g-factor variations between the spins), $\mathbf{S}_i$ ($S_i^z$) is the spin operator ($z$ component) for spin $i$, and $J_{ij}(t)$ is the ideal schedule of exchange pulses between spins $i$ and $j$. Zeeman noise $\delta_i^{\epsilon^z}$ is sampled from the normal distribution $\mathcal{N}(0, \sigma = \sqrt{2}\hbar/T_2^*)$, and unitless exchange error $\delta^{J}_{ij}$ is sampled from $\mathcal{N}(0, \sigma = \Delta^{J})$ for each quasi-static simulation shot. We consider the high magnetic field ($\sim$ mT) limit such that the impact of $\delta_i^{\epsilon^{x,y}}$ is negligible compared to $\delta_i^{\epsilon^z}$.

For each gate two-qubit gate, we simulate the pulse schedule as a unitary evolution of the 6 electron spins, and then compute its effect on the qubit subspace. The two-EO-qubit gate infidelity is calculated as the sum of leakage and unitary error within the qubit subspace. We calculate the average fidelity over 20 quasistatic instances, which we then use in the Stim simulations of the surface code described in the next section. \revision{Figure~{\ref{fig:length_vs_infid}} shows the simulated infidelities of several operations for fixed values of $T_2^*$ and $\Delta^J$, showing that infidelity generally tends to increase with sequence duration as expected. LCCX appears to exhibit little increase in error with increasing sequence length, indicating some degree of built-in noise insensitivity, as observed in \mbox{\cite{weinstein_universal_2023}}.}

\subsection{Teraquop footprint calculation}

We adapt the Stim \cite{gidney_stim_2021} simulation code from \cite{mcewen_data_2023} to calculate the teraquop footprints. After each two-qubit gate in the Stim circuit, we add two-qubit depolarizing noise with probability corresponding to the simulated fidelity of that particular gate. For several choices of code distance, we perform X- and Z-basis logical memory experiments and combine the X and Z error rates to obtain the overall logical failure rate per QEC round. We then project the code distance that would be needed to reach a logical failure rate of $10^{-12}$ and calculate the corresponding number of physical qubits needed to construct one logical qubit (the \emph{teraquop footprint}). This footprint is a function of the simulated two-qubit gate error rates from the previous section.

The results are shown in Figure \ref{fig:tqfps}. The three plots show data for the three different parallelism restrictions in different noise regimes. We sweep ${t_{\text{pulse}}}/{T_2^*}$ on the x axis for three fixed values for $\Delta^J$. Due to the many simplifying assumptions used in this analysis, we do not intend these results to be exact resource estimates, but rather we wish to study the extent to which the different pulse schedules in the same noise regime can affect the quantum error correction overhead.

For the full parallelism case, we see little difference between hardware layouts, except in the regime of very strong $\Delta^{J}$ noise. However, with parallelism restrictions, differences begin to emerge. For NN-restricted parallelism, with $\Delta^{J} = 0.5\%$ and ${t_{\text{pulse}}}/{T_2^*} \approx 0.1\%$, HexTri (green) has a teraquop footprint of around 300 qubits, while DenseHexTri's (yellow) footprint exceeds 10,000 qubits to create a logical qubit of equivalent error rate. HexLinear and GridL would require about 1,000 qubits each.

Overall, these results show that program-level performance cannot be predicted simply by looking at individual pulse sequences for linear or triangular connectivities. It is critical to evaluate pulse sequences in the context in which they will be performed in real applications to draw meaningful conclusions.

\section{Discussion}

\revision{In this work, we described a general method to adapt any exchange-only pulse sequence to some restricted dot connectivity by optimally inserting spin swaps. By providing a way to easily adapt pulse sequences to different connectivities, this tool allows for novel pulse sequences to be designed in a connectivity-agnostic way, potentially enabling the further development of advanced operations like the leakage-controlled CX and CZ gates.}

\revision{This approach is fast for two-EO-qubit (six-spin) systems, enabling us to} generate a complete set of pulse sequences to implement two-qubit gates on exchange-only spin qubits in hundreds of unique connectivities. Compared to linear connectivity, we found a reduction of up to 42.8\% in sequence lengths across all connectivities, which could yield considerable improvements in gate fidelity if taken into account when designing hardware. We introduced the new abstraction of gates with flexible spin permutations, achieving a free additional reduction in average sequence length. 

However, as we saw in the QEC case study, the length of individual pulse sequences does not tell the whole story. Device-level parallelism restrictions can dramatically change the relative performance of different connectivities, and relatively minor differences in pulse sequences can translate to critical differences in resource cost. The exponential error suppression of QEC greatly rewards any minor improvement in physical performance and punishes any inefficiency, so it is crucial to take considerations like this into account when designing scalable architectures for QEC. Pulse sequences are only one part of the operation of a quantum computer, but may be a key part of the make-or-break situation typical of QEC.

We envision several areas of future work to further explore pulse-aware layout design and to improve pulse sequences themselves. Our QEC study involved four handpicked layouts, but a more systematic automated search of the quantum dot array design space is needed, paying more attention to potential restrictions arising from large internal readout components and fabrication constraints. Additionally, incorporating gate fidelity directly into the optimization, rather than relying on pulse count as a proxy, may lead to pulse sequences with better on-chip performance. \revision{For example, dynamical decoupling on exchange-only qubits can be performed by shuffling the spins between dots such that each spin spends equal time in each dot \mbox{\cite{sun_fullpermutation_2024}}; it would be interesting to explore how similar ideas could be incorporated into this swap-adding pulse optimization to improve noise resilience.}

This work is the first step towards pulse-informed, QEC-informed hardware design for EO systems. The design space for quantum dot arrays is extensive, giving the opportunity to tailor a layout and schedule to the specific QEC scheme and hardware constraints at hand. We believe that similar co-design considerations will be critical to enable highly scalable, highly performant quantum computing on exchange-only qubits in silicon.\\

\section*{Data availability}

The complete library of pulse sequences is publicly available at \cite{chadwick_data_2024}.

\section*{Author Contributions}
J.D.C. generated the pulse library, performed simulations, and compiled the manuscript. G.G.G. developed preliminary pulse sequence simulation code and assisted with the theory of EO qubit gates. J.D.C., F.L., and M.T.M. integrated the pulse library into the device control software. F.L. and M.T.M. performed device experiments. F.A.M. assisted with modelling noise in silicon spin qubits. P.P. and J.S.C. suggested dot topologies to consider. P.P. suggested the CXSWAP operation and envisioned uses for the CXSWAP. J.D.C, A.L., and S.P. integrated the pulse library into the Intel Quantum Compiler. A.T.S., G.G.G., N.C.B., A.Y.M., and J.S.C. coordinated the project and provided guidance on direction. All authors revised the manuscript.

\bibliography{2024-eo-pulses}

\begin{figure*}
    \centering
    \includegraphics[width=\linewidth]{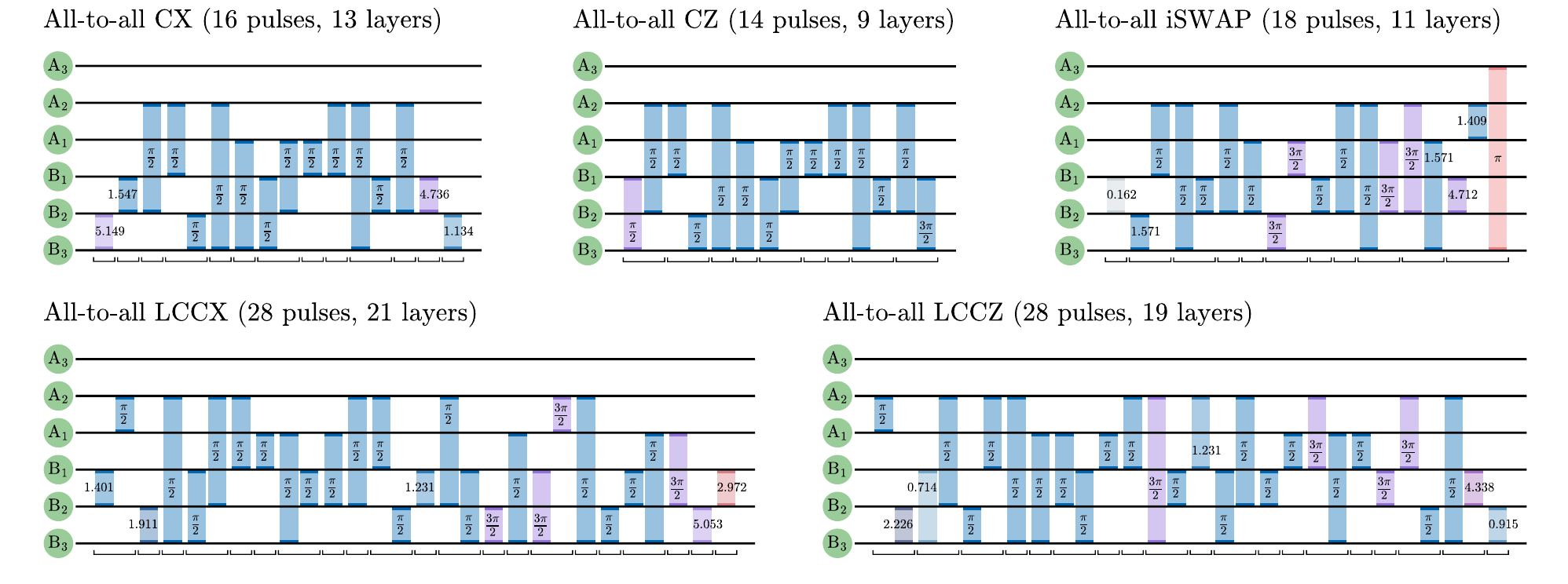}
    \caption{All-to-all sequences used as reference sequences in the optimization algorithm.}
    \label{fig:a2a-pulses}
\end{figure*}

\newpage

\appendix

\section{All-to-all pulse sequences}\label{app:all-to-all}

In Figure \ref{fig:a2a-pulses}, we show the all-to-all pulse sequences for CX, CZ, iSWAP, LCCX, and LCCZ that are used in the optimization to obtain the pulse sequences presented in the main text.

The all-to-all CX pulse sequence was obtained by adding local corrections (the four pulses with non-$\pi/2$ angles) to the sequence from Ref. \cite{setiawan_robust_2014} to yield a full CX operation. The CZ was obtained similarly with different local corrections. The iSWAP operation is locally equivalent to CX + SWAP + single-qubit corrections, so we took the all-to-all CX pulse, added an all-to-all SWAP (three $\pi$ pulses) and the appropriate single-qubit corrections, and then removed/fused exchange pulses to reduce the final sequence length. The all-to-all LCCX and LCCZ were obtained by removing/fusing exchange pulses starting from the sequences presented in \cite{weinstein_universal_2023}.

\end{document}